\documentclass[aps,prl,twocolumn,showpacs,showkeys,amsmath,amssymb]{revtex4}
\usepackage{amsfonts}
\usepackage{amsmath}
\usepackage{graphicx}
\usepackage{subfigure}
\usepackage{dcolumn}
\usepackage{bm}
\usepackage{booktabs}
\usepackage[utf8]{inputenc}

\usepackage{graphicx,graphics,dcolumn,booktabs,bm}
\usepackage{longtable,lscape}
\usepackage{txfonts}
\usepackage{overpic}
\usepackage{amssymb}
\usepackage{indentfirst}
\usepackage{epsfig}
\usepackage{feynmf}   
\usepackage{epstopdf}   
\usepackage{slashed}  
\usepackage{multirow}

\usepackage{ulem}
\usepackage[usenames,dvipsnames]{color}

\makeatletter
\newcommand{\figcaption}{\def\@captype{figure}\caption}
\newcommand{\tabcaption}{\def\@captype{table}\caption}

\newcommand{\Rmnum}[1]{\expandafter\@slowromancap\romannumeral #1@}
\def\hlinewd#1{%
  \noalign{\ifnum0=`}\fi\hrule \@height #1 \futurelet
   \reserved@a\@xhline}
\makeatother

\def\f(s){[(\alpha+\beta)m^2-\alpha\beta s]}

\begin{document}

\title{The possible tetraquark states $cc \bar c \bar c$ observed by the LHCb experiment}

\author{Kuang-Ta Chao}
\email{ktchao@pku.edu.cn}
\author{Shi-Lin Zhu}
\email{zhusl@pku.edu.cn} \affiliation{School of Physics and State
Key Laboratory of Nuclear Physics and Technology, Peking University,
Beijing 100871, China
\\
Center of High Energy Physics, Peking University, Beijing 100871,
China }

\begin{abstract}
We give a brief comment on the possible tetraquark states $cc \bar c
\bar c$ observed by the LHCb experiment.
\end{abstract}

\maketitle


The molecular positronium (sometimes also denoted as di-positronium)
was the multi-electron bound state composed of $e^+e^-e^+e^-$, where
the electromagnetic interaction is the underlying driving force.
Although its existence was predicted as early as in 1947
\cite{1947-Hylleraas-p493-496}, it remained elusive until it was
produced experimentally in 2007 \cite{2007-Cassidy-p195-197}. One
may wonder whether there exists the analogue of the di-positronium
in Quantum Chromodynamics, which is the tetraquark state with the
flavor configurations $q\bar q q \bar q$. Very recently, the LHCb
Collaboration observed distinct structures with the $cc \bar c \bar
c$ (two charm quarks and two anti-charm quarks) in the $J/\psi$-pair
mass spectrum \cite{cccc}. They reported a broad structure ranging
from 6.2 to 6.8 GeV and a narrow structure at around 6.9 GeV with a
global significance of more than $5\sigma$. Let's denote it as
$T_{4c}$.

When Gell-Mann \cite{1964-Gell-Mann-p214-215} and Zweig
\cite{1964-Zweig-p-} proposed the quark model, they also speculated
the possible existence of the multiquark states beyond conventional
mesons and baryons. Later, the low-lying scalar meson nonet below 1
GeV was suggested as possible candidates of the tetraquark states
composed of four light quarks due to their unusual mass ordering
\cite{1977-Jaffe-p281-281,2007-Chen-p94025-94025}. Since the
discovery of X(3872) in 2003, more and more candidates of multiquark
states were reported over the past decades, including dozens of
charmonium-like $XYZ$ states~\cite{pdg} and the hidden-charm
pentaquarks $P_c(4312)$ and $P_c(4450)$
\cite{2015-Aaij-p72001-72001,lhcb2019}.

The hidden-charm pentaquark $P_c$ states and many XYZ states such as
X(3872) and $Z_c(3900)$ lie very close to the di-hadron threshold.
For example, the strong couple-channel effect between the $c\bar c$
core and the $\bar D D^\ast$ continuum may play a pivotal role in
the formation of the X(3872) state \cite{meng}. Without the $\bar D
D^\ast$ components, it will be difficult to explain its isospin
violating decay mode $J/\psi \pi\pi$. Similarly, the $Z_c(3900)$
signal may also arise from the multiple channel scattering.

On the other hand, the $P_c$ states are very good candidates of the
loosely bound $\Sigma_c^{(*)}{\bar D}^{(*)}$ molecular states
\cite{yang1,zou1,zhang1,report,rmp}. The same chiral dynamics
associated with the light quarks is responsible for the existence of
both the deuteron and $P_c$ states. In the framework of the meson
exchange model, the pseudoscalar meson, scalar meson and vector
meson exchange forces provide the attraction. In the chiral
effective field theory, the long-range one-pion, medium-range
di-pion exchange force and contact interactions contribute to the
formation of the hadronic molecules.

Either the light quark degree of freedom or the channel coupling (or
both) is crucial to the existence of the $P_c$ and many XYZ states.
In contrast, the observed $T_{4c}$ structures lie well above the
$\eta_c\eta_c$ threshold (and the narrow structure around 6.9~GeV is
even higher than the $J/\psi J/\psi$ threshold by 700~MeV), and they
do not contain any light quarks. The molecular states composed of a
pair of the doubly-charm baryon and anti-baryon lie above (or
around) 7.2 GeV \cite{menglu}. Therefore, the $T_{4c}$ structures
are unlikely to be the hadronic molecules, which are usually formed
by light meson exchanges with small binding energies. Namely, the
$T_{4c}$ structures do not suffer any complications from the channel
coupling and chiral dynamics. These signals may be good candidates
of the "genuine" compact tetraquark states arising from the
quark-gluon interaction in QCD
\cite{1975-Iwasaki-p492-492,1981-Chao-p317-317,1982-Ader-p2370-2370,1985-Heller-p755-755,2004-Lloyd-p14009-14009,1992-Silvestre-Brac-p2179-2189,1993-Silvestre-Brac-p273-282,2006-Barnea-p54004-54004,plb2017,prd2019,00,0,1,2,3,4,5,6}
.

The $T_{4c}$ structure provides a new platform to investigate the
low-energy dynamics of QCD. The color configurations of the
traditional $q\bar q$ meson and $qqq$ baryon are uniquely determined. However, the
color wave function of di-charm quarks may be ${\bar 3}_c$ or $6_c$.
In other words, there are two possible color configurations ${\bar
3}_c \times 3_c$ and ${\bar 6}_c \times 6_c$. One may wonder which
configuration leads to the $T_{4c}$ structure with lower energy. The
answer is deeply rooted in the confinement mechanism. Moreover,
these two color configurations may mix due to the chromomagnetic
interactions.

In the traditional quark model, in addition to the long-ranged confining force, one generally considers the two-body short-ranged
interaction from the gluon exchange, which follows the similar
formalism in atomic physics. However, the non-Abelian SU(3) gauge
group of QCD differs from the U(1) of QED greatly. There exist the
triple-gluon and quartic-gluon interactions in QCD. Very luckily,
these non-Abelian interactions vanish and do not contribute to the
traditional meson and baryon spectrum due to their unique color
configuration in quark model. However, the situation is very
different for the $T_{4c}$. The color wave function of any three
quarks within the $T_{4c}$ is ${\bar 3}_c$ or $3_c$. Now the genuine
three-body interaction from the triple-gluon or quartic-gluon
interaction does not vanish. This effect has never been investigated
in the literature. Moreover, it is also important to understand the long-ranged
confining force within the $T_{4c}$ states.

In summary, the structures observed by the LHCb experiment in
the $J/\psi$-pair mass spectrum may open up a new testing ground for the
"genuine" compact tetraquark states arising from the quark-gluon
interaction in QCD, aside from the hadronic molecules that are loosely bound by light meson exchanges. This significant signal found by LHCb clearly awaits the
confirmation of other experiments. Determining the quantum numbers of these structures, and finding structures with other charmonium pairs than the $J/\psi$-pair are all very helpful. With more progress of
the $T_{4c}$, $T_{4b}$ and $T_{2c2b}$ investigation in the future,
we may gain new insight on the confinement mechanism and the non-Abelian
low-energy dynamics of QCD.


\section*{Acknowledgments}
This project is supported by the National Natural Science Foundation
of China under Grants 11975033, 11745006.

\end{document}